\documentclass[journal=langd5,manuscript=article]{achemso}

\usepackage{chemformula} 
\usepackage[T1]{fontenc} 

\author{Yaxing Li}
\affiliation{Physics of Fluids group, Department of Science and Technology, Mesa+ Institute for Nanotechnology, Max Planck Center for Complex Fluid Dynamics and 
J. M. Burgers Centre for Fluid Dynamics, University of Twente, P.O. Box 217, 7500 AE Enschede, The Netherlands}
\author{Valentin Salvator}
\affiliation{Physics of Fluids group, Department of Science and Technology, Mesa+ Institute for Nanotechnology, Max Planck Center for Complex Fluid Dynamics and 
J. M. Burgers Centre for Fluid Dynamics, University of Twente, P.O. Box 217, 7500 AE Enschede, The Netherlands}
\author{Herman Wijshoff}
\affiliation[Unknown University]
{Department of Mechanical Engineering, Eindhoven University of Technology, P.O. Box 513, 5600 MB Eindhoven, The Netherlands}
\alsoaffiliation[Second University]
{Oc\'{e} Technologies B.V., P.O. Box 101, 5900 MA Venlo, The Netherlands}
\author{Michel Versluis}
\affiliation{Physics of Fluids group, Department of Science and Technology, Mesa+ Institute for Nanotechnology, Max Planck Center for Complex Fluid Dynamics and 
J. M. Burgers Centre for Fluid Dynamics, University of Twente, P.O. Box 217, 7500 AE Enschede, The Netherlands}
\author{Detlef Lohse}
\affiliation{Physics of Fluids group, Department of Science and Technology, Mesa+ Institute for Nanotechnology, Max Planck Center for Complex Fluid Dynamics and 
J. M. Burgers Centre for Fluid Dynamics, University of Twente, P.O. Box 217, 7500 AE Enschede, The Netherlands}
\alsoaffiliation[Second University]
{Max Planck Institute for Dynamics and Self-Organization, 37077 G\"ottingen, Germany}
\email{d.lohse@utwente.nl}

\title[An \textsf{achemso} demo]
  {Evaporation-induced crystallization of surfactants in sessile multicomponent droplets}

\abbreviations{IR,NMR,UV}

\begin{document}


\begin{abstract}
Surfactants have been widely studied and used in controlling droplet evaporation. In this work, we observe and study the crystallization of sodium dodecyl sulfate (SDS) within an evaporating glycerol-water mixture droplet.  The crystallization is induced by the preferential evaporation of water, which decreases the solubility of SDS in the mixture. As a consequence, the crystals shield the droplet surface and cease the evaporation. The universality of the evaporation characteristics for a range of droplet sizes is revealed by applying a diffusion model, extended by Raoult's law. To describe the nucleation and growth of the crystals, we employ the 2-dimensional crystallization model of Weinberg, \textit{J. Non-Cryst. Solids}, 1991, \textbf{134}, 116. The results of this model compare favorably to our experimental results. Our findings may inspire the community to reconsider the role of high concentration of surfactants in multicomponent evaporation system.
\end{abstract}

\section{Introduction}
Surfactants are widely used to control the evaporation behavior of sessile droplets on a flat substrate~\cite{Sempels2013,Alvaro2016,Kwiecinski2019}. The motivation is driven by various applications in inkjet printing, surface coating and patterning~\cite{park2006control,kong2014}, which mainly aim to optimize the drying rate and the final deposition. The biggest challenge for a controlled uniform coating by droplet evaporation originates from the well-known ``coffee-stain effect"~\cite{deegan1997}. It has been shown that surfactant-induced Marangoni flow can play an essential role to suppress this effect~\cite{Still2012,kim2016controlled}. In these studies, one of the most common ionic surfactants, ``sodium dodecyl sulfate" (SDS)~\cite{Piret2002,Kumar2015,Choudhary2018} is added to the system at small concentration, typically $\leq 1$ wt\%. The surfactants are therefore considered to be always soluble in the system during most of the evaporation lifetime. However, in many practical cases, the relevant liquids contain a high concentration of surfactants; e.g., liquid detergents can contain  surfactant ingredients at up to 40\% by weight. Such high loading of surfactants may lead to undesired effects, such as separation and crystallization. 

Sodium dodecyl sulfate (SDS) may crystallize in liquid solutions upon cooling~\cite{SMITH2004} or upon seeding with 1-dodecanol~\cite{SUMMERTON2016}. On the other hand, selective evaporation of some liquid components with larger volatilities can also lead to phase separation in multicomponent mixtures~\cite{Li2018,Kim2018,Li2019}. Consequently, the nonvolatile surfactant (SDS) is expected to separate from an evaporating liquid system by crystallization due to the preferential evaporation of volatile liquids. Therefore, the wide-usage of SDS in evaporating droplet systems deserves a more detailed explanation of the crystallization behaviour. 

In this work, we study a multicomponent droplet system consisting of a mixture of glycerol, water, and SDS and let it evaporate in ambient air.  SDS is not miscible with pure glycerol, but it does dissolve in glycerol-water mixtures for large enough water concentration ratios. This behavior qualitatively resembles the ternary ``ouzo'' system~\cite{Vitale2003} consisting of water, ethanol, and anise oil, which nucleates in droplets for low enough ethanol concentrations. Tan \textit{et al.}~\cite{Tan2016,Tan2017a} triggered this emulsification threshold by the selective evaporation of ethanol in an evaporating ouzo droplet. Similarly, the varying solubility of SDS in glycerol-water binary systems may also lead to phase separation due to the concentration change caused by the selective evaporation of water alone. In contrast to crystallization by cooling~\cite{SMITH2004,SUMMERTON2016}, here the oversaturation with SDS and the subsequent nucleation and growth of SDS crystals is caused by the preferential evaporation of water at room temperature~\cite{crystal2018}. 

To better understand the evaporation-induced crystallization in the mixture droplet system, two main questions need to be addressed: how does a surfactant-laden mixture droplet evaporate and how to model the crystallization during the evaporation? In this paper we want to answer these questions. A typical snapshot of an evaporating droplet is shown in Fig.~\ref{fgr:capture}, where the two life phases can be distinguished: the evaporation phase and the crystallization phase. The focus of our study is on the dynamics of the evaporation and the kinetics of crystallization, and not on the micro-scale crystal morphology. 

\section{Experimental Methods}

\subsection{Materials and preparation}

The liquid solution was prepared with an initial composition of 78\% (w/w) Milli-Q water (Reference A+, Merck Millipore, 25$^\circ$), 19.6\% (w/w) glycerol (Sigma Aldrich; purity $\geq 98$) and 2.4\% (w/w) sodium dodecyl sulfate (Sigma Aldrich, purity 98\%). The initial concentration of SDS is 13 CMC (critical micelle concentration). Experiments were carried out on a transparent hydrophobic octadecyltrichlorosilane (OTS)-glass substrate~\cite{Peng2014}. The static contact angle of Milli-Q water and glycerol on the substrate are $105^{\circ} \pm 3^{\circ}$ and $90^{\circ} \pm 3^{\circ}$, respectively. The glycerol-water binary droplet with 50\%/50\% (w/w) has a $95^{\circ}$ static contact angle. Prior to each experiment, the samples were cleaned by sonication in an ultrasonic bath of ethanol and subsequently in water, then dried under a flow of nitrogen gas. 

\subsection{Experimental setup}

\begin{figure}[H]
\centering
  \includegraphics[width=0.98\textwidth]{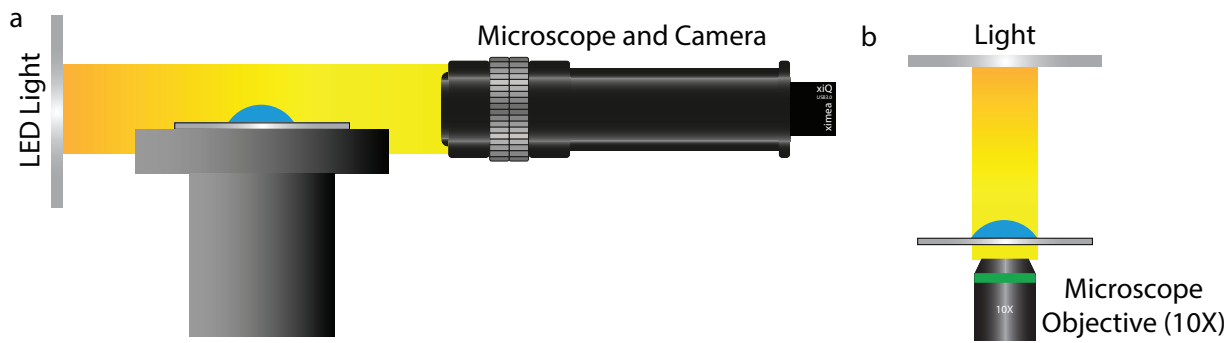}
  \caption{A schematic sketch of the experimental setups. (a) The contact angle device contains a CCD camera with a microscope and a LED light source illuminating the droplet. (b) The droplet is illuminated from above and recorded by a camera equipped with a $10\times$ microscope objective underneath. The whole setup is part of a confocal microscope (Nikon A1 confocal laser microscope system, Nikon Corporation, Tokyo, Japan).}
  \label{fgr:setup}
\end{figure}

We performed two different experiments to separately study the evaporation phase and crystallization phase. To study the evaporation behavior, the droplets were deposited on the substrate by a Hamilton 2 $\mu$L syringe, which was mounted vertically on a computer-controlled motorized pump, that allowed the dispense of droplets of a controlled volume through a needle. We measured the geometry of the deposited droplet by bright-field imaging in side view. The whole process was recorded by an OCA 15 (Dataphysics, Germany) contact angle device (Fig.~\ref{fgr:setup}.a): a CCD camera coupled to a microscope, which was back-illuminated by a LED light from the opposite side of the droplet. For the crystallization study, we observed the droplet in bottom view with a confocal microscope (Fig.~\ref{fgr:setup}.b). By focusing on the layer close to the substrate (at a $\approx 10\ \mu$m height), the dynamic growth of the crystals was visualized in a 2-dimensional view. The experiments were performed at room temperature of 21.4 $\pm$ 1$^{\circ}$C and at relative humidity of 50$\%\pm 5\%$. These parameters were monitored and recorded for each measurement.

\subsection{Imaging analysis}

\begin{figure}[H]
\centering
  \includegraphics[width=0.98\textwidth]{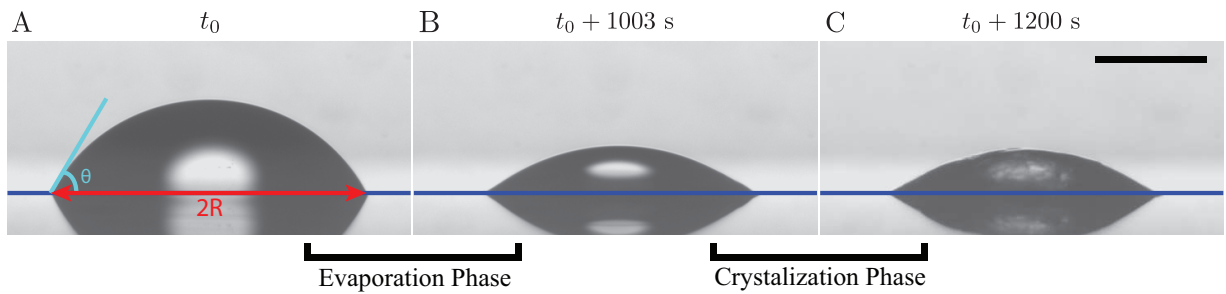}
  \caption{Experimental snapshots of the evaporation and drying process of a typical drop on a flat surface. (A) and (B) show the evaporation phase of the drop: here, the drop retains a spherical cap shape; no crystallization occurs. (C) The final state of the drop: due to the crystallization of SDS in the bulk, the surface of the drop buckles and no longer remains spherical. The crystallization of the SDS shields the surface and brings the evaporation process to an end. The scale bar represents 0.5~mm.}
  \label{fgr:capture}
\end{figure}

For the side-view geometrical measurement, images were analyzed using a custom-made Matlab code to detect the droplet profile with sub-pixel accuracy~\cite{vanderbos2014}. The sizes of all droplets are smaller than the capillary length $\sqrt{\gamma/(\rho g)} \approx 2.7$~mm for the used liquids~\cite{cazabat2010}, where $\gamma \approx 70$~mN/m and $\rho \approx 10^3$ kg/m$^3$ are the surface tension and density of the mixture, and $g = 9.8$~m/s$^2$ is the gravitational acceleration. The detected profile is fitted to a spherical cap during the evaporation phase, which enables us to calculate the volume $V$ of the droplet with footprint radius $R$ and contact angle $\theta$. As shown in Fig.~\ref{fgr:capture}A, the dark blue solid line is the position of the substrate: the spherical shape above it is the sessile droplet, the one underneath is its reflection.

For the documentation of the crystallization process from the bottom view, a manual detection with ImageJ was used to measure the crystallized area at every time instant, see details in Supplementary Materials. 

\section{Experimental results}

\subsection{Evaporation phase}

\begin{figure}[H]
 \centering
 \includegraphics[width=0.98\textwidth]{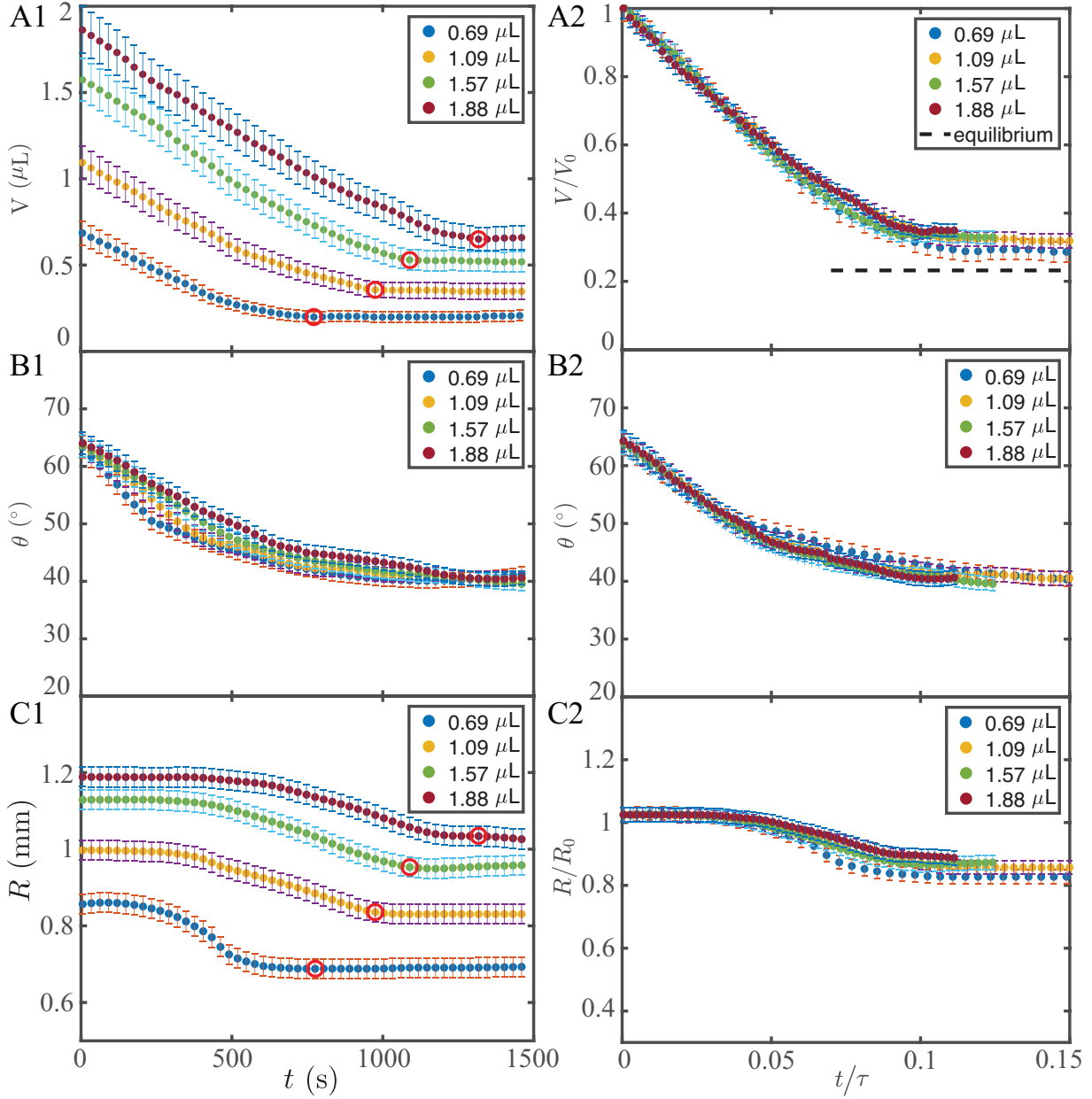}
 \caption{(A1,B1,C1) Measured temporal evolution of the geometrical parameters: volume $V$ (A1), contact angles $\theta$ (B1)and lateral sizes $R$ (C1). The red dots mark the moments when buckling occured. (A2,B2,C2) Same parameters as in experiment, but now non-dimensional and plotted against the scaled time following Eq.~(\ref{eq:scaled_time}). The data collapse clearly shows the universality of the drop evaporation process. (A2) The final volume is controlled by the occurrence of crystalization, rather than by the liquid-vapor equilibrium relation, which is shown by the black dashed line.}
 \label{fgr:parameters}
\end{figure}

\begin{figure}[H]
 \centering
 \includegraphics[width=0.98\textwidth]{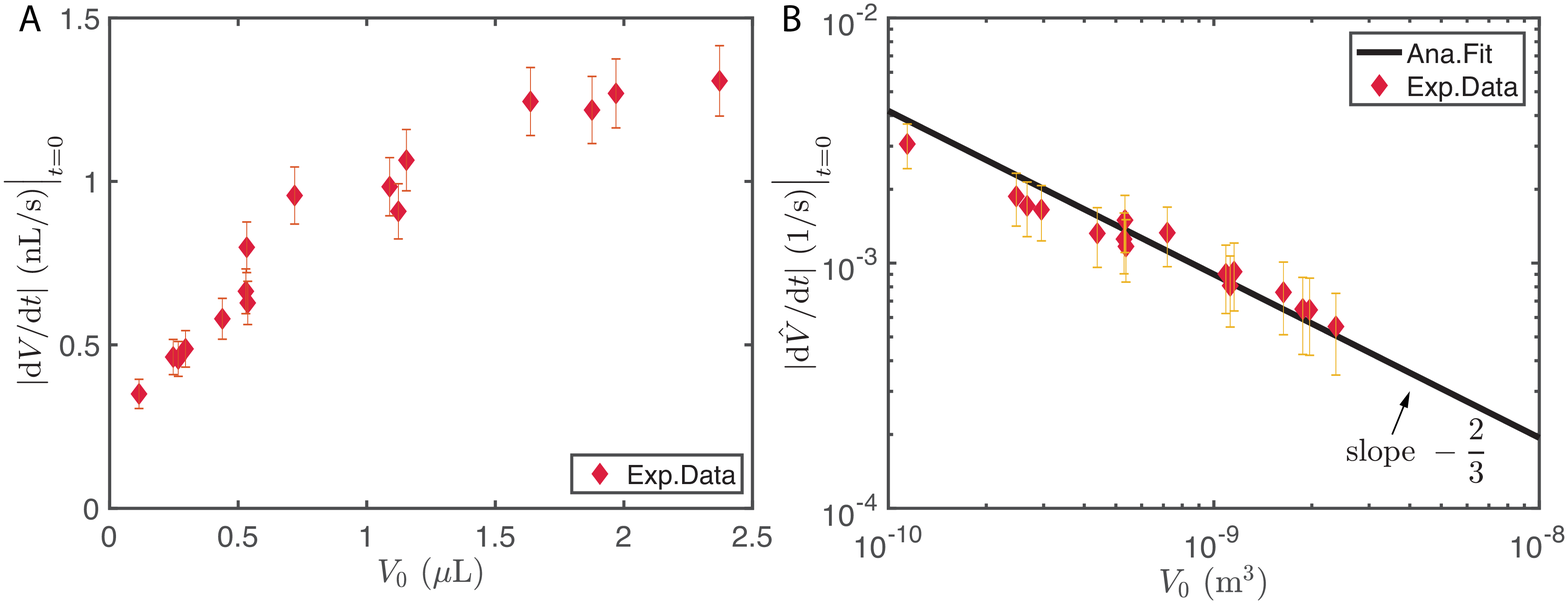}
 \caption{(A) The initial rate of volume loss of the drop varies for different initial volumes. (B) The same data normalized by the initial volume is plotted against the initial volume. The straight line shows the scaling relation with slope -2/3, demonstrating good agreement with the experimental data.}
 \label{fgr:evap_rate}
\end{figure}

The left column of Fig.~\ref{fgr:parameters} displays the temporal evolution of the drop-characterizing geometrical parameters for four droplets with different initial sizes : volume $V$ (A1), contact angle $\theta$ (B1), and footprint radius $R$ (C1). From the plots, it is evident that all the droplets evaporate following the ``stick-slide" mode~\cite{stauber2014,lohse2015rmp}, in which the droplet's footprint radius first remains constant until it reaches a critical contact angle, then the contact line starts to shrink. We only measure the volume until buckling occurs (as marked by the red circles in Fig.~\ref{fgr:parameters}A1, C1), and after that,  the droplet shape deforms and no regular shape is reestablished, which from then on renders accurate volume measurement impossible. Fig.~\ref{fgr:evap_rate}A shows the average evaporation rate of various droplets in the first 30 sec after deposition with initial volumes ranging from 0.12~$\mu$L to 2.40~$\mu$L. The evaporation rate monotonically increases with increasing droplet size, apart from fluctuations due to experimental uncertainties. 

\subsection{Crystallization phase}

\begin{figure}[H]
 \centering
 \includegraphics[width=0.98\textwidth]{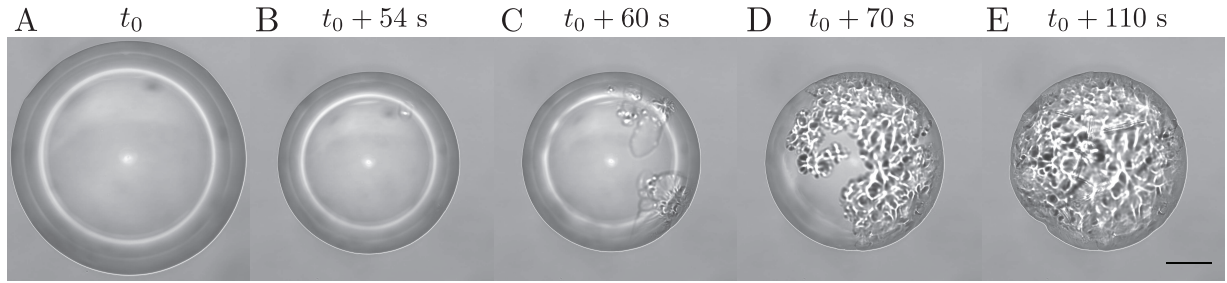}
 \caption{Bottom-view of a complete drop life time. (A) The drop evaporates on the substrate with receding contact line. (B) The first crystal appears near the contact line region. (C) Several crystals nucleate and grow independently. (D) Growing crystals coalesce with neighbouring ones. (E) The crystals cover the whole drop and bring the evaporation to an end. (B to E) The contact line basically remains the same until the final state of the drop, but slightly deforms due to the buckling of the drop surface. The scale bar represents 50~$\mu$m.}
 \label{fgr:crystal_large}
\end{figure}

\begin{figure}[H]
 \centering
 \includegraphics[width=0.98\textwidth]{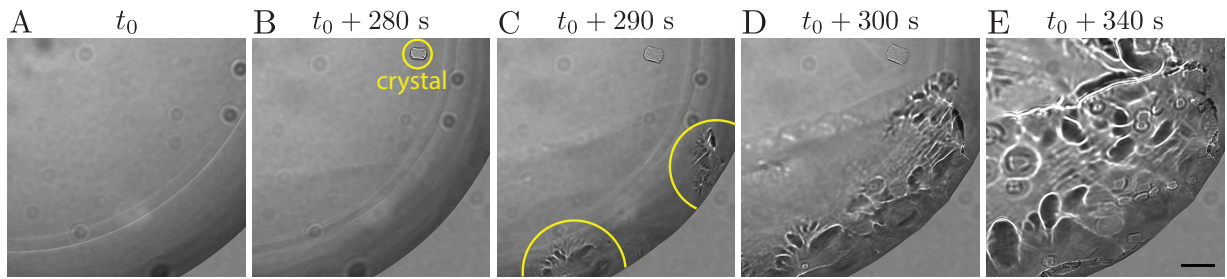}
 \caption{Bottom-view snapshots of the contact region of an evaporating surfactant-binary drop. (A) The moment of deposition of the drop: the drop starts evaporating on the substrate. (B) A small crystal nucleates (yellow circle), floats and grows near the contact line. (C) The crystals heterogenously nucleate at the contact line. (D) The nucleated crystals grow and merge with neighbouring crystals. (E) The crystalized SDS fully occupies the drop and eventually brings the evaporation to an end. The scale bar represents 20~$\mu$m.}
 \label{fgr:crystal}
\end{figure}

\begin{figure}[H]
 \centering
 \includegraphics[width=1\textwidth]{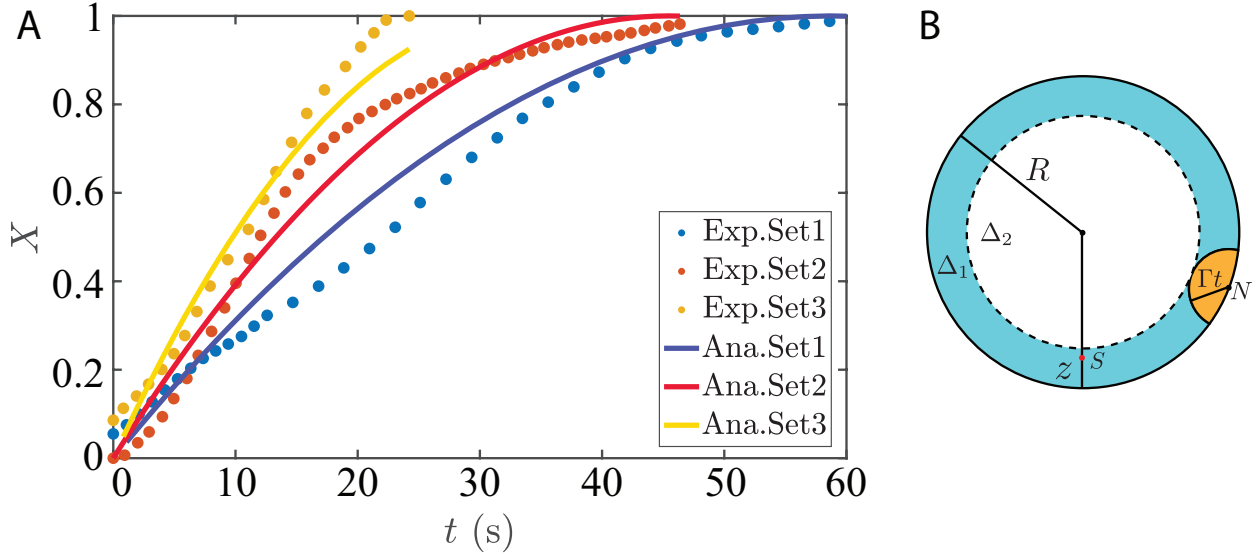}
 \caption{(A) Measurement of the crystallized area fraction of three droplets against time. The analytical results (solid lines) according to Eq.~(\ref{eq:crys_X}) are also shown. (B) The circular regions with areas $\Delta_1$ and $\Delta_2$, represent the areas where nucleation can or cannot occur in time $t$, respectively. The orange region indicates the area transformed at time $t$, due to a nucleation at N. $S$ is an arbitrary point in $\Delta_1$ which has distance $z$ from the boundary. Within time $t$, it must be transformed due to a nucleus on the boundary.}
 \label{fgr:JMAK}
\end{figure}

Figure~\ref{fgr:crystal_large} shows the complete crystallization process of an evaporating surfactant-laden mixture droplet. The droplet starts to evaporate at time $t_0$. At approximately 50 sec, the first crystals appear near the contact line (CL) region. After a few more seconds, several crystals nucleate at the rim. Then they grow and coalesce to form a larger piece and finally occupy the whole bulk of the droplet. Figure~\ref{fgr:crystal} presents a zoomed-in bottom view of the contact region of another evaporating surfactant-laden mixture droplet. Initially, the droplet is transparent with a smooth CL. After evaporating for 280 sec, a crystal nucleates near the CL and floats to the position labeled by the yellow circle. A few seconds later, more crystals nucleate at the rim, slightly deforming the CL. The nucleated crystals grow and coalesce with neighbouring crystals. Eventually, the whole droplet is occupied by the crystals and the CL deforms and is no longer smooth. Figure~\ref{fgr:JMAK}A shows the temporal evolution of the transformed fraction measured in a 2D bottom view for three different droplets. $X$ is the area fraction occupied by crystals and $t$ is the time which has elapsed after the first crystallization had been observed. The area fraction $X$ increases as the growth of crystals at a different rate for each droplet.

\section{Theoretical analysis}

\subsection{Theory of mixture droplet evaporation}

We first study the evaporation characteristics of the surfactant-laden mixture droplet. In general, for a droplet evaporating on a flat surface under ambient conditions and in the absence of any correction, the evaporation is fully controlled by the diffusion of the vapor away from the droplet~\cite{picknett1977,hu2006}: the liquid molecules change their phase and diffuse as vapor molecules into the surrounding air. Popov~\cite{popov2005} derived an analytical solution by using the solution of the equivalent problem of an electric potential around a charged lens-shaped conductor:

\begin{equation} 
\centering
\frac{\text{d}m}{\text{d}t} = -\pi DR(c_{\text{s}}-c_{\infty})f(\theta)
\label{eq:popov}
\end{equation}
with

\begin{equation}
\centering
f(\theta)=\frac{\text{sin}(\theta)}{1+\text{cos}(\theta)}+4\int_{0}^{\infty} \frac{1+\text{cosh}(2\theta\varepsilon)}{\text{sinh}(2\pi\varepsilon)}\text{tanh}[(\pi-\theta)\varepsilon] \text{d}\varepsilon
\end{equation}
with $m$ the droplet mass, $D$ the diffusion coefficient of the droplet liquid vapor in air, $c_s$ the saturated concentration of liquid vapor molecules, and $c_\infty$ the ambient concentration of the liquid vapor far away from the drop.  

For the evaporation of multicomponent droplets, we first employ the method suggested by Brenn~\cite{Brenn2007}, namely considering the total evaporation rate of the mixture droplet as the sum of the evaporation rate of each individual component. In our surfactant-laden glycerol-water droplet, glycerol and SDS are non-volatile under ambient condition\cite{Geballe2010}. Therefore, only the diffusive flux of water contributes to the total evaporation rate. The essential difference between the evaporation of pure droplets and multicomponent droplets is the vapor-liquid equilibrium: the non-volatile component in the system alters the saturated concentration of water vapor at the interface~\cite{Tan2016,Diddens2017a}. Raoult's law~\cite{Raoult1887} is used to calculate the saturated water vapor concentration of the binary system: $c_{w,\text{s}} = X_w c_{w,\text{s}}^{0}$, where $X_w$ is the mole fraction of water at the interface and $c_{w,\text{s}}^{0}$ is the saturated vapor concentration of pure water. However, Raoult's law relies on an idealized solution and as such ignores any interaction between the components. To overcome this limitation, the so-called activity coefficient $\psi$~\cite{chu_prosperetti_2016} was introduced to describe this interaction. In our case, it specifically addresses the interaction between water and the other components: $c_{\text{w,s}} = \psi_{\text{w}}X_\text{w} c_{\text{w,s}}^{0}$. By using the water activity coefficient $\psi_{\text{w}}$~\cite{marcolli2005} in the modified Raoult's law, we obtain a theoretical model to express the evaporation rate for the binary droplet:

\begin{equation}
\centering
\frac{\text{d}m}{\text{d}t}=-\pi DR(\psi_{\text{w}}X_\text{w} c_{\text{w,s}}^{0}-c_{\text{w,}\infty})f(\theta).
\label{eq:normal_popov}
\end{equation}
There is, however, one added complexity in our system: it is difficult to determine the exact $c_{\text{w,s}}$ without knowing the exact mole fraction of water, glycerol, and SDS molecules. Hence we cannot analytically predict the evaporation rate for each time instant. To compare different sets of experimental data, we rescale the measured droplet volume and time, by introducing the non-dimensional volume $\hat{V} = V/V_0$ and time $\hat{t} = t/\tau_c$, with $V(t)$ the measured droplet volume and $V_0$ its initial volume. $\tau_c$ is the characteristic timescale of the diffusive evaporation~\cite{gelderblom2011,lohse2015rmp}, which can also be read from Eq.~(\ref{eq:popov}),

\begin{equation}
 \tau_c = \frac{\rho R_0^2}{D\Delta c}. 
\label{eq:scaled_time}
\end{equation}

Figures~\ref{fgr:parameters}A2,B2,C2 show that the rescaled experimental data for volume $V/V_0$, contact angle $\theta$ and footprint radius $R/R_0$ versus the dimensionless time $t/\tau_c$ follow a universal curve for all measured droplet sizes. The collapse of all the curves demonstrates that regardless of the initial size, the droplets with the same initial composition always follow the same evaporation behavior, with a universal evolution of all geometrical characteristics. Based on this, we can conclude that the variations of not just the geometry but also of the internal composition concentration and distribution are universal, independent of the droplet size. 

We also compare the initial evaporation rate of different initial volumes by introducing the dimensionless volume loss rate $\text{d}\hat{V}/\text{d}t = \text{d}(V/V_0)/\text{d}t$. According to Eq.~(\ref{eq:normal_popov}),  the dimensionless initial evaporation rate is

\begin{equation}
\frac{\text{d}\hat{V}}{\text{d}t}\bigg|_{t=0} \propto \frac{D\Delta cR_0}{\rho V_0} \propto \frac{D\Delta c}{\rho V_0^{2/3}}.
\label{eq:evap_rate}
\end{equation}
Based on the $V_0^{-2/3}$ proportionality of Eq.~(\ref{eq:evap_rate}), we rescale the experimental data of Fig.~\ref{fgr:evap_rate}A, see Fig.~\ref{fgr:evap_rate}B, and plot them on a double logarithmic scale. Indeed, the data follow the scaling law as suggested by Eq.~\ref{eq:evap_rate}, confirming our model assumptions. 

Besides controlling the evaporation rate, the model also yields the terminal state of the evaporation, which is when the saturated water vapor concentration equals the environmental concentration, $c_{w,s} = c_{w,\infty}$. Essentially, the evaporation stops when the active mole fraction of water equals the relative humidity $H$ of the surrounding air, $\psi_w X_w = H$. For the same reason as mentioned above, we only compare the experimental data with the analytical prediction for glycerol-water binary system, ignoring the mole fraction of the surfactant. From the relative humidity $H$ measured in experiment, we can calculate analytically the ``theoretical final volume" $V_t$ (see Supplementary Materials) as

\begin{equation}
V_t = \left(\frac{M_w}{M_g}\frac{H}{\psi_w-H}+\frac{\rho_w}{\rho_g}\right)\left(\frac{1-C_g}{C_g}+\frac{\rho_w}{\rho_g}\right)^{-1}V_0,
\label{eq:final}
\end{equation}
where $M_g = 9.21 \times 10^{-2}$ kg/mol and $M_w = 1.8 \times 10^{-2}$ kg/mol is the molecular mass of glycerol and water, respectively, $\rho_g = 1.226 \times 10^3$ kg/m$^3$ and $\rho_w = 0.997 \times 10^3$ kg/m$^3$ are their liquid densities at room temperature, and $C_g$ is the initial mass concentration of glycerol in each measurement. The final volume of the equilibrium state (dashed line in Fig.~\ref{fgr:parameters}A2) lies below the final volumes of all the droplets, which indicates that the shielding of water by the crystallized interface blocked any further evaporation before the system reached its equilibrium state.

\subsection{Theory of 2-dimensional finite system crystalization}

As it is well-known, the evaporation rate has a singularity at the rim of the droplet, provided the contact angle is smaller than 90$^\circ$~\cite{deegan1997,Tan2016}, which in our system, where the contact angle ranges from 65$^\circ$ to 40$^\circ$ during the evaporation process, indeed is the case. The singularity implies that the water depletes the fastest at the rim, which locally leads to a higher concentration of glycerol at that part. It is therefore also expected that crystal nucleation occurs first near the CL region due to the highest degree of oversaturation of SDS. 

To model the crystallization, we employ a 2-dimensional model which is derived by extending the JMAK formalism~\cite{Avrami1939,Avrami1940,Avrami1941,Johnson1939,Kolmogorov1937} to a finite 2-dimensional system with non-uniform nucleation. Based on the spherical shape of the droplet, the footprint area is circular and the nucleation starts near the contact line. We assume that the crystallization process occurs within a circular region, and nucleation is permitted at $t = 0$ at various points on the perimeter of the area. Figure~\ref{fgr:JMAK}B shows the geometry of the two regions $\Delta_1$ and $\Delta_2$ within a circle with radius $R$ (drop radius): the $\Delta_2$ region is completely free of crystallization, while $\Delta_1$ is partially crystalline. The maximum growth radius of the crystals is given by $\Gamma t$, where $\Gamma$ is the constant growth rate. Weinberg~\cite{WEINBERG1991,WEINBERG1997} proposed an analytical model to describe the fraction $X(t)$ transformed at time $t$, namely

\begin{equation}
X(t) = [1 - (1-\Gamma t/R)^2] X_1(t).
\label{eq:crys_X}
\end{equation}
where $X_1(t)$ represents the fraction which has crystalized in $\Delta_1$. It can be expressed as~\cite{WEINBERG1997}
\begin{equation}
X_{1}(t) = 1- \int_{1-y}^{1} \text{exp} \left[-2P_{1}R \text{cos}^{-1}\left(\frac{1+\Phi^{2}-y^{2}}{2\Phi}\right)\right] \Phi\text{d}\Phi  \times \frac{1}{2}[1-(1-\Phi)^{2}]^{-1},
\label{eq:crys_X1}
\end{equation}
with $\Phi = (R-z)/R$ and $y = \Gamma t/R$. $z$ denotes the distance between an arbitrary point $S$ in the $\Delta_1$ region and the boundary. $P_1$ is the nucleation probability per unit length in region $\Delta_1$. 

We demonstrate that the transformation rate is more sensitive to the growth rate $\Gamma$ rather than to the seeding probability $P_1$, as shown in Supplementary Materials. Here we set $P_1 = 1000~\mu$m$^{-1}$ by assuming a saturated nuclei density. We test this theory for the three cases in Fig.~\ref{fgr:JMAK}A with droplet footprint radius $R_1 = 146\ \mu$m, $R_2 = 110\ \mu$m and $R_3 = 86\ \mu$m. By fitting the theoretical curves to the experimental data, we obtain the growth rate as the fitting parameter: Quite consistently, the results are $\Gamma_1 = 2.48\ \mu$m/s, $\Gamma_2 = 2.42\ \mu$m/s, and $\Gamma_3 = 2.58\ \mu$m/s for the three analyzed cases. From Fig.~\ref{fgr:crystal}C, we estimate the crystal growth rate in the early crystallization stage by measuring the increasing rate of crystal size near the contact line within the yellow circle. We obtain the estimate $\Gamma \approx 25 \pm 5~\mu$m/10~s = 2.5 $\pm$ 0.5~$\mu$m/s, which is comparable to the values $\Gamma_1, \Gamma_2$, and $\Gamma_3$ obtained from our model. Even though we applied a 2D model to a 3D problem, the theoretical predictions show good agreement with experimental data: the reason is that our droplet is relatively flat, with a contact angle of about 40$^\circ$ when crystallization occurs. 

\section{Conclusions and outlook}
In summary, crystallization of sodium dodecyl sulfate induced by selective evaporation in a surfactant-laden glycerol-water mixture droplet is observed during the evaporation process. We studied experimentally the dynamics of evaporation prior to the occurrence of crystal nucleation and the kinetics of crystallization, thereafter. We applied a diffusion model extended by Raoult's law to describe the evaporation characteristics and could reveal a universal evaporation behavior, independent of the size of the droplets. Finally, we applied a 2-dimensional model building on the JMAK nucleation model to describe the kinetics of the crystallization. Thanks to the the low contact angle, this model can successfully describe our experimental data on nucleation.

Surfactants attract significant attention as their ubiquitous role in fluid dynamics of either nature or technology~\cite{manikantan2020}. Our findings clearly show an unexpected consequence of using surfactants in such evaporating systems. This particularly holds for inkjet printing where surfactants are extensively used. As nearly all inks contain various components with different volatilities, the variations of the composition ratio caused by the selective evaporation of more volatile components may lead to the segregation of surfactants in the form of liquid phase separation~\cite{Li2018} or crystallization. Our study may rise the awareness of using surfactants with cautions in such multicomponent systems, which normally involves rich physicochemical processes~\cite{lohse2020}.

Some issues remain open and unexplored. As the temperature can change the CMC of SDS in glycerol-water mixture~\cite{ruiz2008}, does the crystallization behavior also depend on the temperature? How to describe the buckling behavior after the occurrence of crystallization? Another question is on the morphology of the SDS crystals, e.g., is the crystal structure different from the one induced upon cooling? Such questions are of great interest in view of crystal chemistry, and it is worthwhile to further investigate such crystallization behavior from a microscopic perspective in the future.

\begin{acknowledgement}

We thank Shuai Li for valuable suggestions on the manuscript. This work is part of an Industrial Partnership Programme (IPP) of the Netherlands Organization for Scientific Research (NWO). This research programme is co-financed by Canon Production Printing Netherlands B.V., University of Twente and Eindhoven University of Technology. DL gratefully acknowledges support by his ERC-Advanced Grant DDD (project number 740479).

\end{acknowledgement}


\bibliography{achemso-demo}

\end{document}